\documentclass[aps,prb,reprint,twocolumn,superscriptaddress,showpacs]{revtex4-1}
\usepackage{graphicx}
\usepackage{latexsym}
\usepackage{amssymb}
\usepackage{amsmath}
\usepackage{amsfonts}
\usepackage{bm}
\usepackage{multirow}
\usepackage{color}
\usepackage{soul}
\soulregister\ref7
\soulregister\cite7
\soulregister\onlinecite7

\begin{document}

\title{Magnetic correlation between two local spins in a quantum spin Hall insulator}

\author{Ru Zheng}
\email{zhengru@ruc.edu.cn}
\affiliation{Department of Physics, School of Physical Science and Technology, Ningbo University, Ningbo 315211, China}

\author{Rong-Qiang He}
\email{rqhe@ruc.edu.cn}
\affiliation{Department of Physics, Renmin University of China, Beijing 100872, China}

\author{Zhong-Yi Lu}
\email{zlu@ruc.edu.cn}
\affiliation{Department of Physics, Renmin University of China, Beijing 100872, China}


\begin{abstract}
Two spins located at the edge of a quantum spin Hall insulator (QSHI) may interact with each other via indirect spin-exchange interaction mediated by the helical edge states, namely the Ruderman-Kittel-Kasuya-Yosida (RKKY) interaction, which can be measured by the magnetic correlation between the two spins. By means of the newly developed natural orbitals renormalization group (NORG) method, we investigated the magnetic correlation between two Kondo impurities interacting with the helical edge states, based on the Kane-Mele (KM) model defined in a finite zigzag graphene nanoribbon (ZGNR) with spin-orbital coupling (SOC). We find that the SOC effect breaks the symmetry in spatial distribution of the magnetic correlation, leading to anisotropy in the RKKY interaction. Specifically, the total correlation is always ferromagnetic (FM) when the two impurities are located at the same sublattice, while it is always antiferromagnetic (AFM) when at the different sublattices. Meanwhile, the behavior of the in-plane correlation is consistent with that of the total correlation. However, the out-of-plane correlation can be tuned from FM to AFM by manipulating either the Kondo coupling or the interimpurity distance. Furthermore, the magnetic correlation is tunable by the SOC, especially that the out-of-plane correlation can be adjusted from FM to AFM by increasing the strength of SOC. Dynamic properties of the system, represented by the spin-staggered excitation spectrum and the spin-staggered susceptibility at the two impurity sites, are finally explored. It is shown that the spin-staggered susceptibility is larger when the two impurities are located at the different sublattices than at the same sublattice, which is consistent with the behavior of the out-of-plane correlation. On the other hand, our study further demonstrates that the NORG is an effective numerical method for studying the quantum impurity systems.
\end{abstract}


\maketitle

\section{Introduction}
\label{sec:introduction}
The QSHI, namely the 2D topological insulator, has been intensively investigated in recent years after its theoretical prediction~\cite{Kane20052,Bernevig2006} and first discovery in HgTe/CdTe quantum wells~\cite{Konig2007}. The SOC plays an essential role~\cite{Hasan2010,Qi2011} in the QSHI. As a consequence, there is a full insulating gap in the bulk, but there exist one-dimensional gapless conducting edge states with quantized conductance of $G = 2e^2/h$ and opposite spins counterpropagating at each edge, called the helical liquid~\cite{CWu2006}. The time-reversal symmetry (TRS) protects the helical edge states from backscattering, thus they are robust against weak interactions and perturbations preserving the TRS~\cite{Kane20051,CWu2006,Xu2006}. The situation may change when a quantum impurity interacts with the helical edge states, since the backscattering with spin-flip is allowed. The effect of a quantum impurity on the transport properties of the helical edge states has been investigated~\cite{CWu2006,Maciejko2009,Yoichi2011,Joseph2012,Erik2013,Assaad2013,Hu2013}. It is argued that~\cite{Maciejko2009} the conductance of a helical edge state preserves the quantized value $e^2/h$ at zero temperature due to the formation of a Kondo singlet with a complete screening of the impurity spin. On the other hand, the SOC may influence the Kondo effect in the QSHIs~\cite{Zitko2011, Zarea2012,Isaev2012,Kikoin2012,Grap2012,Mastrogiuseppe2014,Wong2016,Sousa2016}. Furthermore, the RKKY interaction between two local spins, which interact with a QSHI, will be mediated by the helical edge states.

The RKKY interaction, mediated by conduction electrons, is an indirect spin-exchange interaction between two local spins. To the second-order perturbation in $J$, the Kondo exchange coupling between two localized spins ${\bf S}_{1,2}$ and conduction electrons, the effective RKKY interaction has the form of
\begin{equation}
H_{\text {RKKY}} = K(R){\bf S}_1 \cdot {\bf S}_2.
\label{eq:RKKY}
\end{equation}
Here the coupling $K(R)$ depends on the distance $R$ between the two spins and $K(R) \propto J^2$ for weak coupling $J$. The physics of a Kondo system with two local spins is determined by the competition between the RKKY interaction and the Kondo effect, which is governed by the ratio of $K(R)$ with respect to the Kondo temperature $T_K \propto e^{-1/{\rho J}}$ with $\rho$ denoting the electronic density of states at the Fermi level. Generally, when $|K(R)| \gg T_K$ (for small $J$), the RKKY interaction dominates over the direct Kondo exchange interaction and the two localized spins will be locked into a singlet (for $K(R)>0$) without the Kondo effect or a screened triplet state (i.e., the two spins align parallel) with weak Kondo effect (for $K(R)<0$)~\cite{Craig2004,Oreg2005,Vavilov2005}. On the other hand, when the coupling $K(R)$ becomes comparable to the Kondo temperature $T_K$, i.e., $K(R) \sim T_K$, a second-order quantum phase transition, controlled by a non-Fermi-liquid fixed point separating the Kondo-screened phase from the interimpurity singlet phase, may occur when the system preserves the particle-hole symmetry~\cite{Jones1988,Jones1989,Jones1995,He2015}.

It has been proposed that controllable RKKY interaction can be used to manipulate the quantum states of local spins, which is very helpful for spintronics as well as quantum computing~\cite{Craig2004,Glazman2004,Oreg2005}. Recently, the RKKY interactions in graphene~\cite{Vozmediano2005,Dugaev2006,Brey2007,Saremi2007,Schaffer2010,Allerdt2015,Allerdt2017} and spin-orbital systems~\cite{Imamura2004,Lai2009,Gao2009,Mross2009,Lee2015,Zare2016,Kurilovich2017,Eickhoff2018,Verma2019} have been intensely investigated. It has been demonstrated that~\cite{Brey2007,Saremi2007} for a honeycomb lattice at half filling, with hopping only between different sublattices, the RKKY interaction is FM for impurities located at the same sublattice and AFM for impurities at the different sublattices. As a comparison, theoretical analysis on the RKKY interaction mediated by the helical edge states, based on the noninteracting low-energy approximation model of the helical edge states and the second-order perturbation theory, shows that the exchange coupling between two local spins is in-plane and noncollinear, and the angle between the two spins depends on the Fermi level of the system~\cite{Gao2009}. In particular, when the Fermi level is near the Dirac point, the exchange coupling becomes a constant and is always AFM~\cite{Gao2009}. This indicates that the helicity of edge states prohibits out-of-plane coupling. Nevertheless, breakdown of this behavior arises in a finite system~\cite{Verma2019}, due to the fact that the helical edge states can come back by traversing the whole edge of the finite system. Considering that the RKKY interaction can be measured by the magnetic correlation between two magnetic impurities, it is thus quite intriguing to investigate the magnetic correlation between two local impurities in a QSHI. Accordingly, the magnetic correlation between two Anderson impurities located at the same sublattice in a graphene nanoribbon with the SOC has been studied by the Quantum Monte Carlo simulation at finite temperatures~\cite{Hu2015}, which shows that the in-plane components of correlations favor ferromagnetism but the out-of-plane correlation can be tuned from ferromagnetism to antiferromagnetism by the SOC. Here with numerical simulation method, we further study the magnetic correlation between two Kondo impurities interacting with the helical edge states, not yet reported in literatures.

In the study, we calculated the magnetic correlation between two Kondo impurities in a QSHI, described by ground state of the KM model~\cite{Kane20052} defined in a finite ZGNR, by using the newly developed NORG method~\cite{He2014}. In particular, the magnetic correlation, including the total correlation as well as its out-of-plane and in-plane components, vs the Kondo coupling and the interimpurity distance were both studied. We further illustrate the influence of relative positions of the two impurities as well as the SOC effect on the magnetic correlation. Additionally, the dynamic properties, represented by the spin-staggered excitation spectrum and the spin-staggered susceptibility at the two impurity sites, were also calculated using the correction vector method~\cite{Kuhner1999,Jeckelmann2002,Schollwock2005}.

This paper is organized as follows. In Sec.~\ref{sec:Model-Method} the KM model and the NORG numerical method are introduced. The energy spectrum of the KM model is shown in Sec.~\ref{sec:EnergySpetrum}. In Secs.~\ref{sec:correlation1} and ~\ref{sec:correlation2}, the magnetic correlation with regard to the Kondo coupling and the interimpurity distance are presented, respectively. Sublattice influence on the magnetic correlation, namely the effect of relative positions of the two impurities, is illustrated in Sec.~\ref{sec:sublattice}. In Sec.~\ref{sec:SOC} the SOC effect on the magnetic correlation is further investigated. Finally, dynamic properties of the system are presented in Sec.~\ref{sec:dynamic}. Section~\ref{sec:summary} gives a short discussion and summary of this work.

\section{Model and numerical method}
\label{sec:Model-Method}
\subsection{Model}
\label{sec:model}
Ground state of the KM model defined in a graphene nanoribbon describes a QSHI with two edge states of opposite spins counterpropagating along each edge, namely, the helical edge states. The KM model can be considered as two copies of the spinless Haldane model~\cite{Haldane1988}, which breaks the TRS. Thus the KM model preserves the TRS. In addition, the helical edge states correspond to the noninteracting limit $K_L = 1$ of a helical Luttinger liquid with $K_L$ representing the Luttinger parameter. In experiment, since the KM model was proposed to describe the quantum spin Hall effect in graphene, extensive strategies have been proposed and developed to enhance the SOC in graphene employing interface or intercalation or doping~\cite{Dedkov2008,Weeks2011,Marchenko2012,Hu2012}.

\begin{figure}[htbp!]
\centering
\includegraphics[width=1.0\columnwidth]{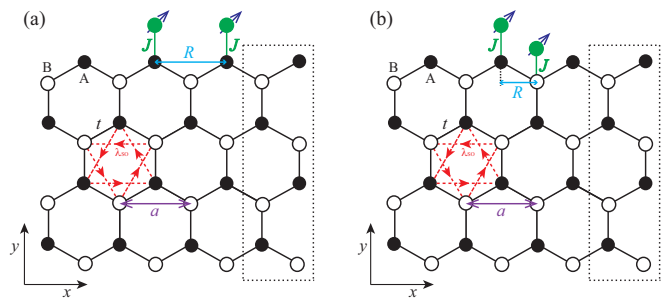}
\caption{\label{fig:Model}(color online) Sketches of the KM model with two Kondo impurities located at the top edge. The ZGNR is periodic (open) along the $x$ $(y)$ direction with length $N_x=4$ and width $N_y=4$. The width $N_y$ of a nanoribbon is defined by the number of zigzag lines. The unit cell of the ZGNR is shown as the dotted black rectangle. The black filled (open) circles denote sublattice A (B) of the nanoribbon. The black lines denote the NN hopping connecting sites of the different sublattices AB. The SOC term connecting sites of the same sublattice AA or BB is denoted as the red dashed arrows, and its sign is associated with $\nu_{ij}$. The two spin-$\frac{1}{2}$ impurities are marked by the filled green circles and they interact directly with the sites of sublattice A or B at the top edge. The distance between the impurities is given as $R \equiv |r_{12}^x|$ with ${\bf{r}}_{12} = {\bf{r}}_1-{\bf{r}}_2$. The two impurities are located at (a) the same sublattice AA with $R=a$ and (b) the different sublattices AB with $R=a/2$. Here $a$ is the lattice constant. $J>0$ is the AFM Kondo coupling strength.}
\end{figure}

Here we consider the Hamiltonian $H_{\text{KM}}$ of the KM model as follows,
\begin{equation}
H_{\text KM} = -t\sum\limits_{\langle ij\rangle\sigma}c_{i\sigma }^ {\dagger}{c_{j\sigma }} +
              i\lambda_{\text{SO}}\sum\limits_{\langle\langle ij\rangle\rangle \alpha\beta}\nu_{ij}c_{i,\alpha}^{\dagger}\sigma_{\alpha\beta}^zc_{j,\beta}.
\label{eq:KM}
\end{equation}
Here $c_{i\sigma}^{\dagger}$ creates an electron at site $i$ with spin component $\sigma=\uparrow,\downarrow$. $\langle ij\rangle$ denotes the nearest-neighbor (NN) hopping and $t$ is the corresponding hopping parameter. $\langle\langle ij\rangle\rangle$ marks the next-nearest-neighbor (NNN) hopping with a complex hopping integral. $\lambda_{\text{SO}}$ represents the strength of the SOC with $\lambda_{\text{SO}}=0.1t$ in our calculations without additional statement. For a smaller value of $\lambda_{\text{SO}}=0.03t$, our calculations give the same physics as well. The parameter $\nu_{ij} = -\nu_{ji} = \pm 1$ depends on the orientation of the two NN bonds that an electron hops from site $j$ to $i$, namely $\nu_{ij} = +1$ if the electron turns left in the hopping from site $j$ to $i$ and $\nu_{ij} = -1$ if it turns right, as shown in Fig.~\ref{fig:Model}. In the SOC part $H_{\text{SO}}$, $\sigma_{\alpha\beta}^z$ is the $z$ Pauli matrix which further distinguishes the spin-up and spin-down states with opposite NNN hopping amplitude.

Due to the fact that the edge states are localized at the edges and exponentially decay into the bulk, we set the two Kondo impurities only located at the top edge of a ZGNR, as presented in Fig.~\ref{fig:Model}. The total Hamiltonian of the system is given by $H = H_{\text{KM}} + H_{\text{Kondo}}$ with
\begin{equation}
H_{\text{Kondo}}=J\sum\limits_{i=1,2} {\bf {S}}_i \cdot {\bf{s}}({\bf{r}}_i),
\label{eq:Model}
\end{equation}
where $H_{\text{Kondo}}$ describes the Kondo exchange interactions between the two spin-$\frac{1}{2}$ impurities and the electrons in the helical edge states. Here each local spin ${\bf {S}}_i$ interacts directly with the conduction electron spin-density ${\bf{s}}(r_i)=\frac{1}{2}\sum_{\alpha\beta}c_{i\alpha}^ {\dagger}{{\bf{\sigma}}_{\alpha\beta}}{c_{i\beta}}$ located at position ${\bf{r}}_i$ with AFM Kondo coupling $J>0$, where ${\bf{\sigma}}$ represents the vector of Pauli matrices. Previous works~\cite{Andrew2017b,Zheng2018} demonstrate that the edge states along the top edge reside mainly in sublattice A. We thus keep one of the impurities coupled to a site of sublattice A, while the other is coupled to another site of either sublattice A or B, i.e., the two impurities are located at the same sublattice AA (Fig.~\ref{fig:Model}(a)) or the different sublattices AB (Fig.~\ref{fig:Model}(b)).

As we see, the total Hamiltonian $H$ breaks the spin-rotation ${\rm SU}(2)$ symmetry by the SOC term with $[{\bf{S}}, H] \neq 0$, but it preserves both charge ${\rm U}(1)_{\rm charge}$ symmetry and spin ${\rm U}(1)_{\rm spin}$ symmetry. Thus the $z$-component of the total spin is still conserved with $[S_{\rm total}^z, H] = 0$. Furthermore, the whole system preserves the TRS.

In the calculations, we set the NN hopping parameter $t$ as the energy unit with $t=1.0$ and kept half-filling for the conduction band. All the calculations were carried out in the ground state subspace of $S_{\rm total}^z=0$. Here the system size was always $L = N_x\times N_y = 28 \times 6$ without additional statement, $N_x$ ($N_y$) denoting the length (width) of the ZGNR. The periodic (open) boundary condition was adopted along the $x$ $(y)$ direction, as schematically shown in Fig.~\ref{fig:Model}.

\subsection{Numerical method}
\label{sec:method}
We employed the NORG approach (see Ref.~\onlinecite{He2014} for details), a newly developed numerical many-body approach without perturbation, to study the magnetic correlation between the two Kondo impurities. It has been demonstrated that the NORG method works efficiently on quantum impurity models in the whole coupling regime~\cite{He2014,He2015,Zheng2018,Zheng2020,Zheng2021}. Moreover, the NORG method preserves the whole geometric information of a lattice and its effectiveness is independent of any topological structure of a lattice.

Generally, the realization of the NORG method essentially involves a representation transformation from the site representation into the natural orbitals representation through iterative orbital rotations. As a result, the NORG method works in the Hilbert space constructed from a set of natural orbitals, which correspond to the eigenvectors of the single-particle density matrix (or the correlation matrix)~\cite{PerOlov1955,Luo2010,Zgid2012,Lin2013,Lu2014,He2014,Fishman2015,Lu2019} defined by $D_{ij}  = \langle\Psi| c_{i}^{\dagger}c_{j}|\Psi\rangle$ with $|\Psi\rangle$ a normalized many-body wave function of the system and $c_{i}^{\dagger}$ the creation operator in the site representation.

More specifically, one performs the representation transformation from site representation into natural orbitals representation by $d_{m}^{\dagger}=\sum_{i=1}^NU_{mi}^{\dagger}c_{i}^{\dagger}$, here $d_{m}^{\dagger}$ represents the corresponding creator in the natural orbitals representation and $U$ is an $N \times N$ unitary matrix diagonalizing the single-particle density matrix $D=U\Lambda U^{\dagger}$ with $\Lambda$ denoting a diagonal matrix and $N$ the system size. In practice, to efficiently realize the NORG approach, only the bath orbitals are transformed into a natural orbitals representation, namely, we rotate only the orbitals of the bath. Therefore, by using the NORG method, we can solve hundreds of noninteracting bath sites with any topological structures, while the computational cost is about $O(N_{\text {bath}}^3)$ with $N_{\text {bath}}$ denoting the number of bath sites.

As a detailed example, after the representation transformation involved in the NORG method, the Kondo interaction $H_{\text{Kondo}}=J\sum_{i=1,2} {\bf {S}}_i \cdot {\bf{s}}({\bf{r}}_i)$ (Eq.~(\ref{eq:Model})) is given by the following forms
\begin{equation}
\begin{split}
{H_{\text{Kondo}}=}&\frac{J}{2}\sum\limits_{i=1,2}\Big\{ \sum\limits_{mn}U_{im}U_{in}^{\dagger}S_i^z(d_{m\uparrow }^ {\dagger}d_{n\uparrow }-d_{m\downarrow}^{\dagger}d_{n\downarrow}) \\
                 & -\sum\limits_{mn}U_{im}U_{in}^{\dagger}c_{i\uparrow }^ {\dagger}d_{n\uparrow }d_{m\downarrow }^ {\dagger}c_{i\downarrow }  \\
                 & -\sum\limits_{mn}U_{im}U_{in}^{\dagger}d_{m\uparrow }^ {\dagger}c_{i\uparrow }c_{i\downarrow }^ {\dagger}d_{n\downarrow } \Big\}
\label{eq:HamiltonianKondo1}
\end{split}
\end{equation}
with $c_{i\uparrow }^ {\dagger}(c_{i\uparrow })$ denoting the creation (annihilation) operator with spin up at the $i$th impurity site.

\section{Numerical results}
\label{sec:Results}
\subsection{Band structure}
\label{sec:EnergySpetrum}
\begin{figure}[htp!]
\centering
\includegraphics[width=1.0\columnwidth]{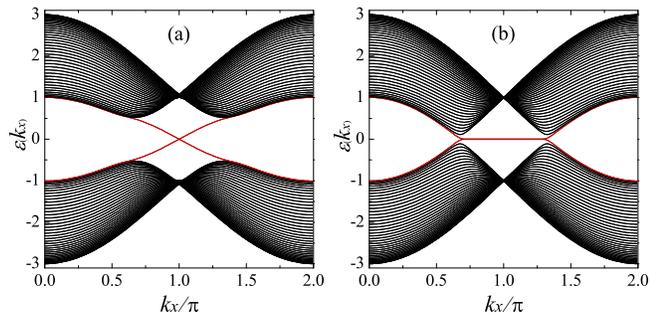}
\caption{\label{fig:EnergySpectra}(color online) Energy spectrum of the KM model in a ZGNR with SOC (a) $\lambda_{\text{SO}} = 0.1$ and (b) $\lambda_{\text{SO}} = 0$, respectively. The edge-state bands in (a) cross with each other at the Fermi level $\varepsilon(k_x=\pi) = 0$, and each band is doubly degenerate. For $\lambda_{\text{SO}} = 0$ in (b), the flat bands related with the edge states emerge. The size of the nanoribbon adopted in the calculation, as sketched in Fig.~\ref{fig:Model}, is $L = N_x\times N_y$ with length $N_x=256$ and width $N_y=40$.}
\end{figure}

Figure~\ref{fig:EnergySpectra}(a) shows the energy spectrum of the KM model in a ZGNR with SOC $\lambda_{\text{SO}} = 0.1$. The two edge-state bands in the spectrum in Fig.~\ref{fig:EnergySpectra}(a) cross with each other at the Fermi level $\varepsilon(k_x=\pi) = 0$, and each band is doubly degenerate according to the Kramers degeneracy. Meanwhile, we also plot the energy spectrum for the vanishing SOC $\lambda_{\text{SO}} = 0$ in Fig.~\ref{fig:EnergySpectra}(b). As expected, the flat bands related with the localized edge states emerge when the SOC vanishes.

\begin{figure}[htp!]
\centering
\includegraphics[width=1.0\columnwidth]{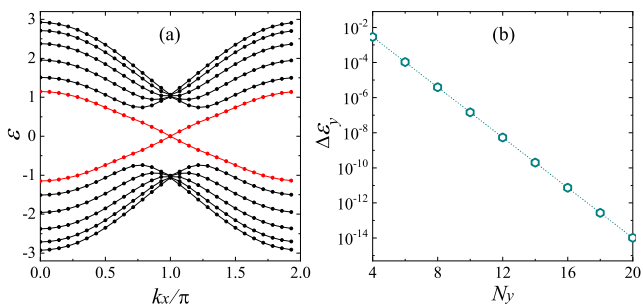}
\caption{\label{fig:EnergySpectraFinite}(color online) (a) Energy spectrum of the KM model in a ZGNR of size $L = N_x\times N_y = 28 \times 6$ and (b) the finite-size energy gap $\Delta \varepsilon_y$ at $k_x = \pi$ with respect to the width $N_y$ for $\lambda_{\text{SO}} = 0.1$. $\Delta \varepsilon_y$ decays exponentially with the nanoribbon width $N_y$ as expected. The nanoribbon length $N_x$ is fixed to $N_x = 28$ in (b).}
\end{figure}

On the other hand, in realistic systems, the edge states decay exponentially into the bulk. Consequently, for a ZGNR with a finite width, the helical edge states coming from the two edges can couple together with a finite overlap to produce a small energy gap at $k_x = \pi$, destroying the QSH effect. In Fig.~\ref{fig:EnergySpectraFinite} we show the energy spectrum of the KM model in a ZGNR of size $L = N_x\times N_y = 28 \times 6$, as well as the finite-size gap $\Delta \varepsilon_y$ at $k_x = \pi$ with respect to the nanoribbon width $N_y$. As we see in Fig.~\ref{fig:EnergySpectraFinite}(b), the energy gap $\Delta \varepsilon_y$ decays exponentially with the nanoribbon width $N_y$ as expected. Specifically, the finite-size energy gap $\Delta \varepsilon_y(N_y=6) \approx 10^{-4}$, indicating that the ground state of the KM model defined in a ribbon with width $Ny=6$ is appropriate to simulate the helical edge states.

\subsection{Magnetic correlation vs the Kondo coupling}
\label{sec:correlation1}
In order to explore the magnetic correlation between the two impurities, which measures the RKKY interaction mediated by the edge states, we first calculate the static spin-spin correlation $\langle{\bf {S}}_1 \cdot {\bf {S}}_2\rangle$ vs the Kondo coupling $J$. In the large Kondo coupling regime, we expect that the Kondo effect dominates over the RKKY interaction, with the two local spins being screened separately. This leads to the interimpurity correlation $\langle{\bf {S}}_1 \cdot {\bf {S}}_2\rangle \to 0$, indicating that the two impurities decouple from each other. As the Kondo coupling $J$ decreases, the behavior of the interimpurity correlation $\langle{\bf {S}}_1 \cdot {\bf {S}}_2\rangle$ is intricate and expected to depend on the relative positions of the two magnetic impurities. In the following calculations in this section, the interimpurity distance is fixed to $R=a$ when the two impurities are located at the same sublattice and that is fixed to $R=a/2$ when at the different sublattices.

\begin{figure}[htp!]
\centering
\includegraphics[width=1.0\columnwidth]{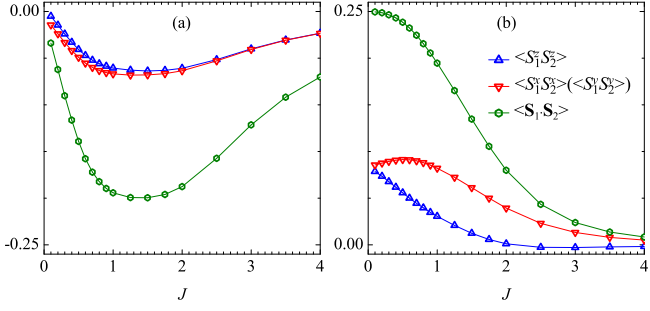}
\caption{\label{fig:CorrelatorJ}(color online) Spin-spin correlation $\langle{\bf {S}}_1 \cdot {\bf {S}}_2\rangle$ and the corresponding components along $z$ direction $\langle S_1^zS_2^z\rangle$ and $x(y)$ direction $\langle S_1^xS_2^x\rangle(\langle S_1^yS_2^y\rangle)$ as functions of Kondo coupling $J$ for fixed interimpurity distances (a) $R = a/2$ and (b) $R = a$, respectively. $\langle{\bf {S}}_1 \cdot {\bf {S}}_2\rangle \to 0$ in the large $J$ regime, indicating that the two impurities are screened separately and thus decouple from each other. In comparison, in the weak $J$ regime, the behavior of the magnetic correlation depends on the interimpurity distance $R$.}
\end{figure}

We plot the calculated $\langle{\bf {S}}_1 \cdot {\bf {S}}_2\rangle$ as a function of $J$ for $R = a/2$ and $R = a$ in Figs.~\ref{fig:CorrelatorJ}(a) and \ref{fig:CorrelatorJ}(b), respectively. As we see, $\langle{\bf {S}}_1 \cdot {\bf {S}}_2\rangle < 0$ for $R = a/2$, which demonstrates that the total correlation between the two impurities located at the different sublattices is AFM. Moreover, in the weak coupling limit $J \to 0$, the spin correlation $\langle{\bf {S}}_1 \cdot {\bf {S}}_2\rangle \to 0$ for $R = a/2$, meaning that the two impurities decouple from each other. Hence when the two impurities are located at the different sublattices, the Kondo effect overwhelms the RKKY interaction in the weak coupling regime. In contrast, $\langle{\bf {S}}_1 \cdot {\bf {S}}_2\rangle > 0$ for $R = a$, indicating that the magnetic correlation between the two impurities at the same sublattice is FM. Furthermore, $\langle{\bf {S}}_1 \cdot {\bf {S}}_2\rangle \to \frac{1}{4}$ when $J \to 0$ for $R = a$. This means that the two spins are locked into a triplet in the weak coupling limit. The resulting triplet may be then screened in a weak two-stage Kondo effect~\cite{Jayaprakash1981}. As the Kondo coupling $J$ increases, the Kondo effect tends to dominate over the RKKY interaction. In consequence, the correlation $\langle{\bf {S}}_1 \cdot {\bf {S}}_2\rangle$ decays smoothly to 0 in the large $J$ regime, as shown in Fig.~\ref{fig:CorrelatorJ}(b).

On the other hand, the SOC effect, which breaks the spin-rotation ${\rm SU}(2)$ symmetry of the total Hamiltonian $H$, should influence the symmetry in spatial distribution of the magnetic correlation. We thus study the components along $x$, $y$, and $z$ directions of the total correlation, namely the in-plane components and out-of-plane component, respectively. When the two impurities are located at the same sublattice, as we see from Fig.~\ref{fig:CorrelatorJ}(b) for distance $R = a$, the symmetry in spatial distribution is preserved with isotropic correlations in the weak coupling $J$ limit, i.e., $\langle S_1^zS_2^z \rangle = \langle S_1^xS_2^x\rangle (\langle S_1^yS_2^y\rangle)$ when $J \to 0$. This symmetry is then broken as $J$ increases, due to the SOC effect. In contrast, when the two impurities are at the different sublattices, this symmetry is slightly broken and tends to recover in the large $J$ regime, as shown in Fig.~\ref{fig:CorrelatorJ}(a) that $\langle S_1^zS_2^z \rangle = \langle S_1^xS_2^x\rangle (\langle S_1^yS_2^y\rangle)$ when $J$ is large.

Moreover, we find that the behavior of the in-plane correlation is consistent with that of the total correlation. Specifically, the in-plane components $\langle S_1^xS_2^x\rangle$ and $\langle S_1^yS_2^y\rangle$ are always AFM when the two impurities are located at the different sublattices, while they are always FM at the same sublattice. For the out-of-plane correlation $\langle S_1^zS_2^z \rangle$, as we can see from Fig.~\ref{fig:CorrelatorJ}(a), it is always AFM when the two impurities are located at the different sublattices with fixed $R = a/2$. However, when the two impurities are at the same sublattice with fixed $R = a$, it changes from FM to weakly but not negligibly AFM as the Kondo coupling $J$ increases. As a consequence, the out-of-plane correlation can be tuned from FM to AFM by manipulating the Kondo coupling $J$.

\subsection{Magnetic correlation vs interimpurity distance}
\label{sec:correlation2}
\begin{figure}[htp!]
\centering
\includegraphics[width=1.0\columnwidth]{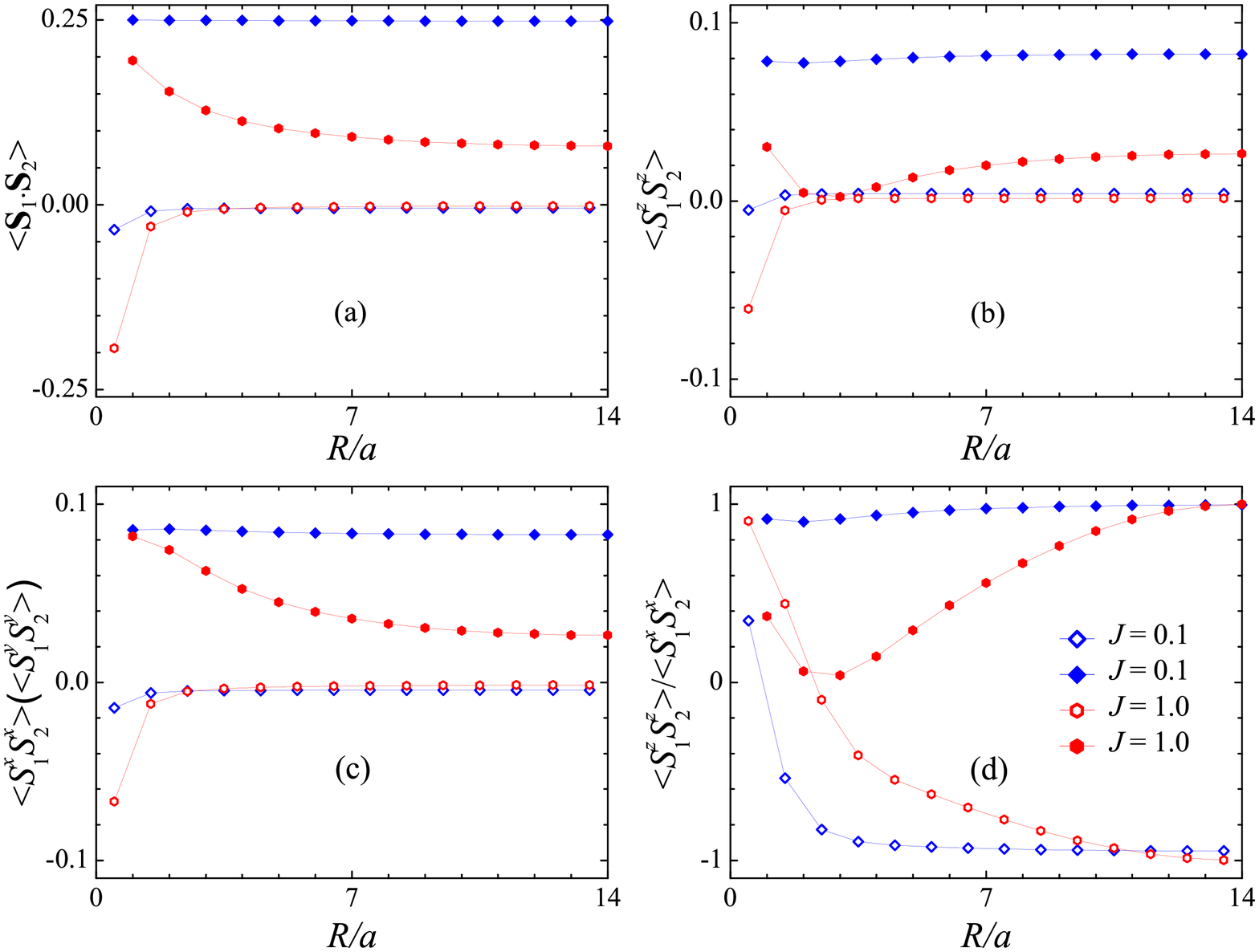}
\caption{\label{fig:CorrDistance}(color online) (a) Spin-spin correlation $\langle{\bf {S}}_1 \cdot {\bf {S}}_2\rangle$ and its corresponding components along (b) $z$ direction $\langle S_1^zS_2^z\rangle$ (out-of-plane correlation) as well as (c) $x$ $(y)$ direction $\langle S_1^xS_2^x\rangle$ $(\langle S_1^yS_2^y\rangle)$ (in-plane components) in regard of the interimpurity distance $R$, respectively. The ratio $\langle S_1^zS_2^z\rangle/\langle S_1^xS_2^x\rangle$, which exhibits the effect of SOC on the symmetry in spatial distribution of the magnetic correlation, is presented in (d). The numerical results are denoted by open (solid) symbols for interimpurity distance $R$ that is of half-integral (integral) multiple of the lattice constant, i.e., when the two magnetic impurities are located at the different sublattices (same sublattice).}
\end{figure}

Considering that the RKKY interaction is sensitive to the distance, we then study the magnetic correlation in regard of the interimpurity distance with fixed Kondo couplings $J=0.1$ and $J=1.0$, namely in the weak and intermediate Kondo coupling regimes. Corresponding numerical results are shown in Fig.~\ref{fig:CorrDistance}.

In Fig.~\ref{fig:CorrDistance}(a) we present the total spin-spin correlation $\langle{\bf {S}}_1 \cdot {\bf {S}}_2\rangle$ as a function of the interimpurity distance $R$. In both cases of $J = 0.1$ and $1.0$, when $R$ is of half-integral multiple of the lattice constant, the correlation $\langle{\bf {S}}_1 \cdot {\bf {S}}_2\rangle <0$ and its magnitude decays with $R$ but does not change its sign, as presented by open symbols in Fig.~\ref{fig:CorrDistance}(a). This indicates that when the two impurities are at the different sublattices, the total correlation between the two impurities are always AFM. While when $R$ is of integral multiple of the lattice constant, the correlation $\langle{\bf {S}}_1 \cdot {\bf {S}}_2\rangle$ is always positive, meaning FM correlation. Specifically, $\langle{\bf {S}}_1 \cdot {\bf {S}}_2\rangle \approx \frac{1}{4}$ for $J = 0.1$, demonstrating that the two impurities are locked into a triplet. Meanwhile, $\langle{\bf {S}}_1 \cdot {\bf {S}}_2\rangle$ decays smoothly as the distance $R$ increases for $J = 1.0$. As a result, when the two impurities are located at the same sublattice, the total magnetic correlation are always FM.

Figure~\ref{fig:CorrDistance}(b) shows the out-of-plane correlation, i.e., $z$ component $\langle S_1^zS_2^z\rangle$ of the spin-spin correlation. When the two impurities are located at the different sublattices, namely $R$ is of half-integral multiple of the lattice constant, $\langle S_1^zS_2^z\rangle <0$ at short distance and then it changes the sign as $R$ increases, meaning that the out-of-plane correlation turns from AFM to FM. In the case of $R$ being integral multiple of the lattice constant, i.e., the two impurities are located at the same sublattice, the out-of-plane correlation is always FM with $\langle S_1^zS_2^z\rangle >0$. As a comparison, $\langle S_1^zS_2^z\rangle$ remains nearly unchanged with the interimpurity distance $R$ for $J = 0.1$, while it decays within short distance and increases afterwards as $R$ further increases for $J = 1.0$. Hence, the out-of-plane magnetic correlation can be adjusted by manipulating the interimpurity distance, especially when the two impurities are located at the different sublattices.

As shown in Fig.~\ref{fig:CorrDistance}(c), the in-plane magnetic correlation, namely $x$ $(y)$ component $\langle S_1^xS_2^x\rangle$ $(\langle S_1^yS_2^y\rangle)$ of the total correlation, is always AFM when the two impurities are located at the different sublattices, while it is always FM when at the same sublattice. As we see, the behavior of in-plane components is consistent with that of the total correlation, implying that the behavior of the total magnetic correlation is mainly determined by that of the in-plane components.

The ratio $\langle S_1^zS_2^z\rangle/\langle S_1^xS_2^x\rangle$, which exhibits the effect of SOC on the symmetry in spatial distribution of the magnetic correlation, is presented in Fig.~\ref{fig:CorrDistance}(d). We see that when the two impurities are located at the different sublattices, the SOC always breaks the symmetry in spatial distribution with the ratio $\langle S_1^zS_2^z\rangle/\langle S_1^xS_2^x\rangle <1$ and $\langle S_1^zS_2^z\rangle/\langle S_1^xS_2^x\rangle$ then decays to $-1$ at long distance $R$. On the other hand, when the two impurities are located at the same sublattice for $J = 0.1$, the spatial isotropy is nearly preserved with $\langle S_1^zS_2^z\rangle/\langle S_1^xS_2^x\rangle \approx 1$. However, the symmetry in spatial distribution is broken for $J=1.0$ at short distance, which tends to recover at very long distance afterwards. Thus, at very long distance $R$ when the two impurities are located at the same sublattice, the effect of SOC on the symmetry in spatial distribution of the magnetic correlation vanishes with $\langle S_1^zS_2^z\rangle/\langle S_1^xS_2^x\rangle = 1$.

\subsection{Sublattice influence on the magnetic correlation}
\label{sec:sublattice}
It has been shown above that when the two impurities are located at the same sublattice, the behavior of magnetic correlation is distinct from that when at the different sublattices. We attribute the difference to the fact that the edge states along the top edge reside mainly in sublattice A~\cite{Andrew2017b,Zheng2018}, namely the local density of states (LDOS) $\rho_A$ at the Fermi energy at sublattice A is relatively larger than $\rho_B$ at sublattice B, leading to different effective couplings between the impurities and the edge states $J^{\text{A}}_{\text{eff}}=\rho_{\text{A}} J$ and $J^{\text{B}}_{\text{eff}}=\rho_{\text{B}} J$ with $J^{\text{A}}_{\text{eff}} \gg J^{\text{B}}_{\text{eff}}$. Hence, two different characteristic scales emerge when $T_K^{\text{A}} \gg T_K^{\text{B}}$ with $T_K \propto J_{\text{eff}}$ in the system. On the other hand, in a finite system with the finite-size gap $\Delta$ at the Fermi energy, the finite-size effect may modify the Kondo physics.

Therefore when an impurity is located at sublattice B, for weak Kondo coupling $J$ with $T_K^{\text{B}} \ll \Delta$, the magnetic moment may be underscreened or even completely decoupled from the conduction electrons. Consequently, in the weak coupling limit, when the two impurities are located at the different sublattices, the one located at sublattice B may decouple from the system while the other at sublattice A is fully screened, leading to the interimpurity correlation $\langle{\bf {S}}_1 \cdot {\bf {S}}_2\rangle \to 0$ when $J \to 0$ with the two impurities decoupling from each other, as shown in Fig.~\ref{fig:CorrelatorJ}(a). Moreover, a minimum point appears in Fig.~\ref{fig:CorrelatorJ}(a), which corresponds to the point where $T_K^{\text{B}}$ is of the same order of magnitude as the finite-size gap $\Delta$, namely $T_K^{\text{B}} \sim \Delta$. Our further numerical results (not shown) indicate that the coupling $J$ at the minimum point is pushed to smaller values as the length $N_x$ increases, since the finite-size gap $\Delta$ decreases.

As supplement, we further study the single impurity case, i.e., there is only one Kondo impurity coupled with the edge states at the top edge. Our numerical results (calculated in the $S_{\rm total}^z=\frac{1}{2}$ subspace) show that $\langle S^z_{\text {imp},\text{A}}\rangle=0$ when the impurity is located at sublattice A, indicating that no free local moment at the impurity site can be polarized and hence the impurity spin is completely screened by the conduction electrons, consistent with our previous work~\cite{Zheng2018}. In contrast, when the impurity is located at sublattice B, the spin polarization $\langle S^z_{\text {imp},\text{B}}\rangle \ne 0$ in the weak coupling regime (for example $J \leq 1.0$), meaning that the impurity may be completely decoupled from the conduction electrons when $J \to 0$. As the Kondo coupling increases (for example $J = 1.2$), the impurity spin polarization tends to vanish with $\langle S^z_{\text {imp},\text{B}}\rangle = 0$. Hence when the impurity is located at sublattice B, the local moment can only be fully screened by the conduction electrons for sufficiently large Kondo couplings.

\begin{figure}[htp!]
\centering
\includegraphics[width=0.6\columnwidth]{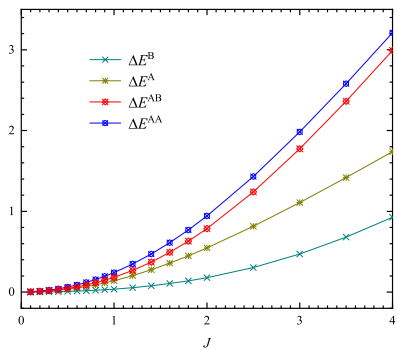}
\caption{\label{fig:KondoEnergy}(color online) The ground-state energy fall after coupling one impurity or two impurities to the topological insulator. $\Delta E^{\text{A}(\text{B})}$ denotes the energy fall when the impurity is located at sublattice A (B), and $\Delta E^{\text{AA}(\text{AB})}$ represents that when two impurities are located at the same sublattice (the different sublattices) with interimpurity distance $R =a$ ($R = a/2$). All calculations are carried out in ZGNRs of size $L = N_x\times N_y = 28 \times 6$ with SOC $\lambda_{\text{SO}} = 0.1$.}
\end{figure}

We proceed to calculate the ground-state energy fall after coupling the impurity to the topological insulator, which is defined as
\begin{equation}
\Delta E^{\text{A}(\text{B})} = E^{\text{A}(\text{B})}_0(J)-E^{\text{A}(\text{B})}_0(J=0),
\label{eq:KondoEnergy1}
\end{equation}
where $\Delta E^{\text{A}(\text{B})}$ denotes the ground-state energy fall when the impurity is located at sublattice A (B). If the impurity is perfectly screened, $\Delta E$ can be identified as an estimate of the Kondo temperature $T_K$ or the energy needed to break the Kondo singlet. As expected, $\Delta E$ increases with the Kondo coupling $J$ and $\Delta E^{\text{A}} > \Delta E^{\text{B}}$, as plotted in Fig.~\ref{fig:KondoEnergy}. In comparison, we also present the ground-state energy fall after coupling two Kondo impurities defined as
\begin{equation}
\Delta E^{\text{AA}(\text{AB})} = E^{\text{AA}(\text{AB})}_0(J)-E^{\text{AA}(\text{AB})}_0(J=0),
\label{eq:KondoEnergy2}
\end{equation}
here $\Delta E^{\text{AA}(\text{AB})}$ represents the ground-state energy fall when the impurities are located at the same sublattice (the different sublattices) with interimpurity distance $R =a$ ($R = a/2$) shown in Fig.~\ref{fig:Model}. Our numerical results, plotted in Fig.~\ref{fig:KondoEnergy}, show that the ground-state energy fall $\Delta E^{\text{AA}} > \Delta E^{\text{AB}}$, in accordance with the results in the single impurity case.

\subsection{SOC effect on the magnetic correlation}
\label{sec:SOC}
\begin{figure}[htp!]
\centering
\includegraphics[width=1.0\columnwidth]{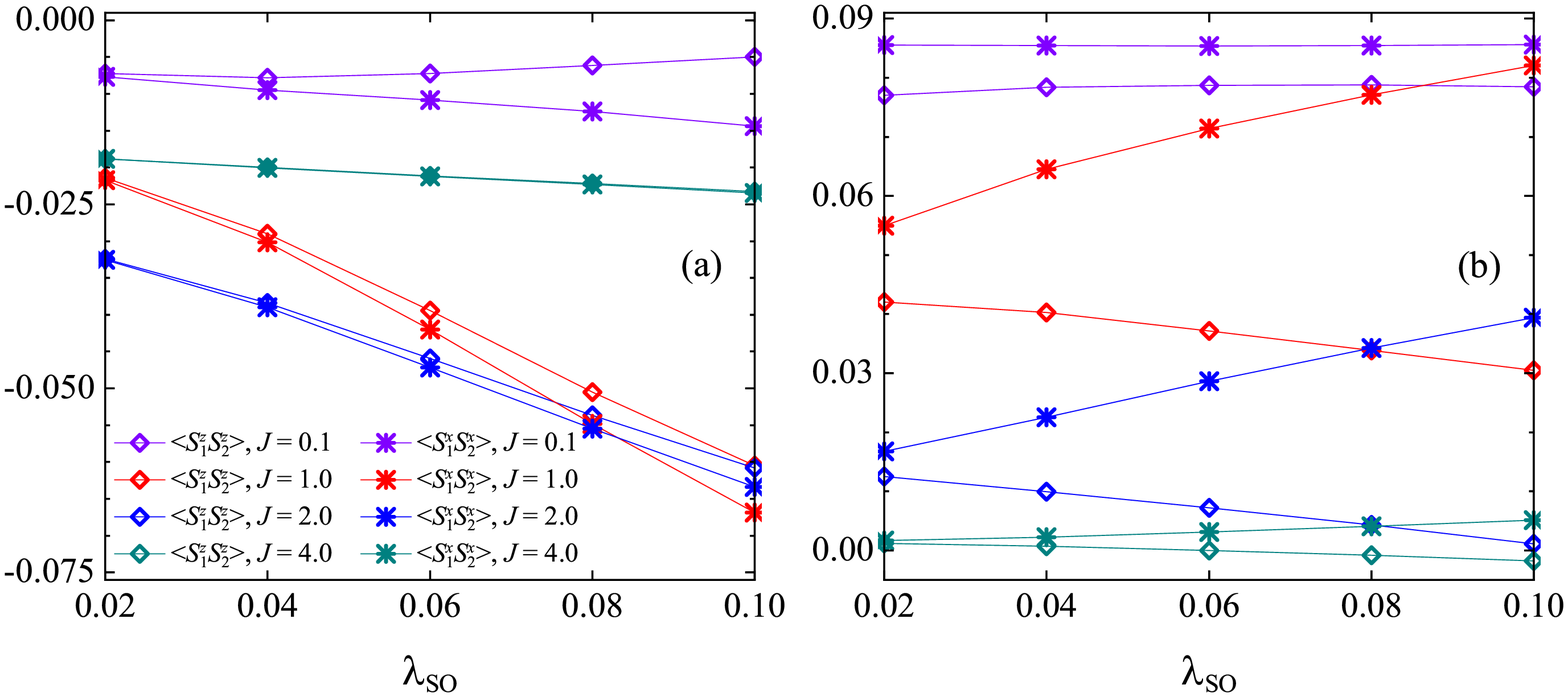}
\caption{\label{fig:SOC}(color online) The magnetic correlation, including the in-plane and out-of-plane correlations, between the Kondo impurities with respect to the strength of SOC $\lambda_{\text{SO}}$ when the impurities are located at (a) the different sublattices with interimpurity distance $R = a/2$ and (b) the same sublattice with $R = a$ in various Kondo coupling regimes. The length (width) of ZGNR used is $N_x=28$ ($N_y=6$).}
\end{figure}

Since the Kondo physics depends drastically on the DOS of the conduction electrons surrounding the magnetic impurities, the RKKY interaction may be modified by the DOS of the free electrons. For the KM model defined in a ZGNR, the LDOS at the Fermi energy at sublattices A and B along the top edge are determined by the strength of SOC $\lambda_{\text{SO}}$. At sublattice A, the LDOS of the edge states associated with the flat bands with the vanishing SOC $\lambda_{\text{SO}}=0$ displays a sharp peak, while the peak is then suppressed and becomes smooth as the $\lambda_{\text{SO}}$ increases with the edge states being broadened. So we expect that the effective coupling between the impurity at sublattice A and conduction electrons will be weaken by the SOC effect. On the other hand, at sublattice B, the LDOS at the Fermi level displays a small but finite value for a nonvanishing $\lambda_{\text{SO}} \neq 0$ and it almost does not vary with the $\lambda_{\text{SO}}$. Thus we propose that the RKKY interaction may be adjusted by the SOC effect. In order to this end, we next study the effect of SOC on the magnetic correlation between the impurities. Numerical results with interimpurity distances $R = a/2$ and $R = a$ are depicted in Figs.~\ref{fig:SOC}(a) and ~\ref{fig:SOC}(b), respectively.

It has been seen in Fig.~\ref{fig:SOC} that the magnitude of the in-plane correlation $\langle S_1^xS_2^x\rangle$ $(=\langle S_1^yS_2^y\rangle)$ increases with $\lambda_{\text{SO}}$, regardless of the relative positions of the impurities. Meanwhile, in the weak Kondo coupling regime (for example $J=0.1$), $\langle S_1^xS_2^x\rangle$ almost does not vary with $\lambda_{\text{SO}}$ when the impurity are located at the same sublattice with the interimpurity distance $R=a$. As comparison, the out-of-plane correlation $\langle S_1^zS_2^z\rangle$ for the interimpurity distance $R=a/2$ behaves distinctly from that for $R=a$. In the case of $R=a/2$, the magnitude of $\langle S_1^zS_2^z\rangle$ increases with $\lambda_{\text{SO}}$ except in the weak coupling regime, where it decreases with $\lambda_{\text{SO}}$. For the interimpurity distance $R=a$, $\langle S_1^zS_2^z\rangle$ declines with $\lambda_{\text{SO}}$, and it turns to negative from positive in the strong coupling regime (for example $J=4.0$).

As a result, the magnetic correlation is tunable by the strength of SOC. Specifically, the in-plane correlation is enhanced by the SOC, except that in the weak coupling regime when the impurity are located at the same sublattice, where it almost does not vary. As for the out-of-plane correlation, its behavior depends on the relative positions of the impurities. When the impurities are located at the different sublattices, the out-of-plane correlation is generally enhanced by the SOC, whereas it is suppressed slightly in the weak coupling regime. In comparison, when the impurities are at the same sublattice, it is overall suppressed as the SOC increases (it is nearly unchanged in the weak coupling regime), and further tuned from FM to AFM by increasing the SOC in the strong coupling regime.

\subsection{Dynamic properties}
\label{sec:dynamic}
In this section, we explore the dynamic properties of the system, represented by the spin-staggered excitation spectrum and spin-staggered susceptibility at the two impurity sites. We study the Green's function in the Lehmann representation defined at the two impurity sites, which has the following form as
\begin{equation}
G_{\text{st}}(\omega) = \langle0|(S_1^z-S_2^z)\frac{1}{\omega+i\eta-H+E_0}(S_1^z-S_2^z)|0\rangle.
\label{eq:Greenfunction}
\end{equation}
Here $|0\rangle$ and $E_0$ denote the ground state and ground-state energy, respectively. The parameter $\eta \to 0$ stands for a Lorentzian broadening factor. The Green's function for a given frequency $\omega+i\eta$ is calculated via the correction vector method.

We first introduce the correction vector method concisely. Consider the following general Green's function $G(A,z)$ in a system with Hamiltonian $H$
\begin{equation}
G(A,z) = \langle0|A^{\dagger}\frac{1}{z-H}A|0\rangle,
\label{eq:Greenfunction}
\end{equation}
where $A$ is the applied operator in our system and $z=\omega+i\eta$. To do the correction vector method, we introduce the first Lanczos vector $|A\rangle=A|0\rangle$ and the correction vector $|x(z)\rangle$ with
\begin{equation}
|x(z)\rangle=\frac{1}{z-H}|A\rangle.
\label{eq:CorrectionVector}
\end{equation}
{We then split the correction vector $|x(z)\rangle$ into real and imaginary part $|x(z)\rangle=|x^r(z)\rangle+i|x^i(z)\rangle$. As a result, the equation for the correction vector Eq.~(\ref{eq:CorrectionVector}) is split into real and imaginary parts $|x^r(z)\rangle$ and $|x^i(z)\rangle$, respectively. The imaginary part  $|x^i(z)\rangle$ is obtained by solving the following equation
\begin{equation}
((H-\omega)^2+\eta^2)|x^i(z)\rangle=-\eta|A\rangle
\end{equation}
using the conjugate gradient method. Furthermore, the real part of the correction vector $|x^r(z)\rangle$ is calculated directly by
\begin{equation}
|x^r(z)\rangle=-\frac{1}{\eta}(H-\omega)|x^i(z)\rangle.
\end{equation}
Finally, the Green's function can be obtained by $G(A,z)=\langle A|x(z)\rangle$.

The spin-staggered excitation spectrum $\chi_{\text{st}}(\omega)=-\frac{1}{\pi}{\text{Im}}G_{\text{st}}(\omega)$ at the two impurity sites is first explored. The behavior of spin-staggered excitation is expected to depend on both the Kondo couplings and the relative positions of the two magnetic impurities, as that of the interimpurity magnetic correlation. As presented in Fig.~\ref{fig:Excitation}, for the weak Kondo coupling ($J = 0.1$), the spin-staggered excitation is concentrated at the point of $\omega=0$ and decays with $\omega$ increasing. In contrast, in the large Kondo coupling regime, for example $J = 4.0$, the spin-staggered excitation spectrum $\chi_{\text{st}}$ tends to vanish, especially when the two impurities are located at the same sublattice ($R = a$), meaning that there is no spin-staggered excitation at the two impurity sites. Here the two impurities are screened separately and decouple from each other for a large Kondo coupling $J$, indicated by the spin-spin correlation $\langle{\bf {S}}_1 \cdot {\bf {S}}_2\rangle \to 0$. As a comparison, for the intermediate Kondo couplings, the spin-staggered excitation is enhanced as $\omega \to 0$ when the two impurities are located at the different sublattices ($R = a/2$), while it is suppressed when at the same sublattice.

\begin{figure}[htp!]
\centering
\includegraphics[width=1.0\columnwidth]{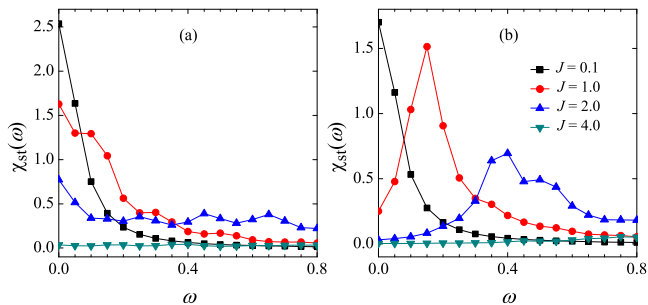}
\caption{\label{fig:Excitation}(color online) Spin-staggered excitation spectrum $\chi_{\text{st}}(\omega)$ at the two impurity sites for the interimpurity distances (a) $R = a/2$ and (b) $R = a$, respectively. The length (width) of ZGNR used is $N_x=16$ ($N_y=6$) with the Lorentzian broadening factor $\eta=t/N_x$.}
\end{figure}

\begin{figure}[htp!]
\centering
\includegraphics[width=0.7\columnwidth]{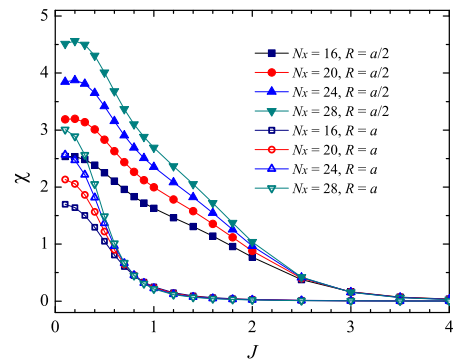}
\caption{\label{fig:Susceptibility}(color online) Spin-staggered susceptibility $\chi$, calculated in the ZGNRs of different length $N_x$ with fixed width $N_y=6$ and the Lorentzian broadening factor $\eta=t/N_x$, as a function of the Kondo coupling $J$ for the interimpurity distances $R = a/2$ and $R = a$, respectively. $\chi$ is larger when the two impurities are located at the different sublattices with $R = a/2$ than at the same sublattice with $R = a$.}
\end{figure}

The spin-staggered susceptibility $\chi$ is then obtained by $\chi=\chi_{\text{st}}(\omega = 0)$. Numerical results calculated in the ZGNRs of different length $N_x$ with fixed width $N_y=6$ are plotted in Fig.~\ref{fig:Susceptibility}. We find that $\chi$ is larger when the two impurities are located at the different sublattices ($R = a/2$) than at the same sublattice ($R = a$). This is consistent with the behavior of the out-of-plane correlation $\langle S_1^zS_2^z \rangle$ shown in Fig.~\ref{fig:CorrelatorJ}, namely, it is always AFM when the two impurities are located at the different sublattices while it changes from FM to weakly AFM with increasing the Kondo coupling $J$ when at the same sublattice. In the large Kondo coupling regime, as expected, influence of the relative positions of the two magnetic impurities on the staggered susceptibility $\chi$ tends to vanish, namely, $\chi \to 0$ when $J$ is large, consistent with the interimpurity correlation $\langle{\bf {S}}_1 \cdot {\bf {S}}_2\rangle \to 0$.

\section{discussion and summary}
\label{sec:summary}
In a QSHI, due to the SOC, one-dimensional gapless conducting edge states with opposite spins counterpropagate at each edge, called the helical edge states. Even though the TRS protects the helical edge states from backscattering, it allows the backscattering accompanied by a spin-flip scattering when quantum impurities interact with the helical edge states. Therefore, two local spins in a QSHI may interact with each other via the RKKY interaction, mediated by the helical edge states. In contrast to the isotropic RKKY interaction in normal metal, the helicity of edge states leads to vanishing out-of-plane RKKY interaction for the spins along an edge in a QSHI, whereas breakdown of this behavior occurs in a finite system. Moreover, due to the spin-momentum locking, spin current is realized in the helical edge states. This indicates that this RKKY interaction is actually an exchange interaction mediated by spin current in a QSHI system. In practical applications, controllable RKKY interaction can be used to manipulate the quantum states of local spins, which is helpful for the spintronics and quantum-information processing. On the other hand, the RKKY interaction can be measured by the magnetic correlation between two local spins. Thus, it is of great importance to investigate the magnetic correlation between two local impurities in a QSHI.

In summary, employing the newly developed NORG method, we investigate the magnetic correlation between two Kondo impurities in a QSHI, based on the KM model defined in a finite ZGNR. We find that the SOC effect breaks the symmetry in spatial distribution of the magnetic correlation, leading to anisotropy in the RKKY interaction. Specifically, the total correlation and its in-plane components are always FM when the two impurities are located at the same sublattice, while they are always AFM when at the different sublattices. However, the out-of-plane component can be tuned from FM to AFM by manipulating either the Kondo coupling or the interimpurity distance. Moreover, the magnetic correlation is tunable by the SOC effect, especially that the out-of-plane correlation can be adjusted from FM to AFM by increasing the SOC when the impurities are located at the same sublattice.

Regarding the different behaviors of the magnetic correlation associated with the relative positions of the impurities, it is attributed to the fact that the edge states along the top edge reside mainly in sublattice A. This means that the LDOS at the Fermi energy at sublattice A is larger than that at sublattice B, resulting in a larger effective coupling between the impurity located at sublattice A and the conduction electrons. On the other hand, the LDOS at the Fermi energy along the top edge is influenced by the strength of SOC. At sublattice A, as the strength of SOC increases, the LDOS is suppressed and then becomes constant with the edge states being broadened. In contrast, at sublattice B, the LDOS displays a small but finite value at the Fermi energy for a nonvanishing SOC and almost does not vary with the SOC. In consequence, the interimpurity RKKY interaction as well as the magnetic correlation is thus tunable by the SOC.

Additionally, dynamic properties of the system, represented by the spin-staggered excitation spectrum and the spin-staggered susceptibility at the two impurity sites, are finally explored. It is illustrated that the spin-staggered susceptibility is larger when the two impurities are located at the different sublattices than at the same sublattice, which is consistent with the behavior of the out-of-plane correlation.

On the other hand, our results further demonstrate that the NORG, whose effectiveness is independent of any lattice structures or topology of a system, is an effective numerical method for studying the quantum impurity problems. Our investigation will promote further theoretical studies on the Kondo effect or the quantum phase transitions in the topological systems by using quantum many-body numerical methods.

\begin{acknowledgments}

This work was supported by National Natural Science Foundation of China (Grants No. 11934020 and No. 11874421). Computational resources were provided by Physical Laboratory of High Performance Computing at RUC.

\end{acknowledgments}

\bibliography{CorrQSHI}

\end{document}